# Allowing Software Developers to Debug HLS Hardware

Jeffrey Goeders and Steven J.E. Wilton
Department of Electrical and Computer Engineering
University of British Columbia, Vancouver, Canada
{jgoeders,stevew}@ece.ubc.ca

*Abstract*—High-Level Synthesis (HLS) is emerging as a mainstream design methodology, allowing software designers to enjoy the benefits of a hardware implementation. Significant work has led to effective compilers that produce high-quality hardware designs from software specifications. However, in order to fully benefit from the promise of HLS, a complete ecosystem that provides the ability to analyze, debug, and optimize designs is essential. This ecosystem has to be accessible to software designers. This is challenging, since software developers view their designs very differently than how they are physically implemented on-chip. Rather than individual sequential lines of code, the implementation consists of gates operating in parallel across multiple clock cycles. In this paper, we report on our efforts to create an ecosystem that allows software designers to debug HLS-generated circuits in a familiar manner. We have implemented our ideas in a debug framework that will be included in the next release of the popular LegUp high-level synthesis tool.

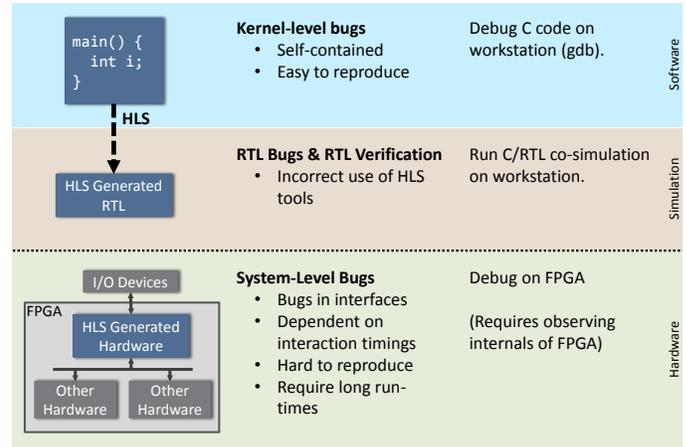

Fig. 1. Classification of bugs in an HLS system

## I. Introduction

High-level synthesis (HLS) allows a program written in a software language (eg. C), to be automatically synthesized into a hardware circuit. HLS is quickly gaining popularity, particularly for FPGA programmable platforms, where it enables their use as compute accelerators alongside traditional processors [1], [2]. Both Altera and Xilinx have invested heavily in this technology, and we anticipate that HLS-based techniques may become the *dominant* design entry method for FPGAs in the future. Intel's recent announcement of their acquisition of Altera further emphasizes the shift of FPGAs from their glue-logic roots to a general-purpose algorithm acceleration platform. This shift will further increase the appetite for the fast turn-around design times and increased accessibility promised by HLS.

In order for HLS to deliver its promised benefits, a compiler is not enough. A complete ecosystem that provides the ability to *analyze, debug, and optimize* designs is essential. To be useful, this ecosystem has to be *accessible to software developers*. Software developers think of their systems in terms of sequential execution of instructions with limited explicit parallelism; this is in contrast to the actual FPGA implementation which consists of interconnected dataflow components operating in parallel across multiple clock cycles. As we will discuss in this paper, this disparity creates a chasm, that if not bridged, will significantly limit the effectiveness of analysis, debug, and optimization; this will, in turn, limit the suitability of HLS in designing most real systems.

Any debugging and optimization ecosystem is likely to be associated with some amount of on-chip instrumentation. In our previous publications, we have focused on the optimization of this instrumentation (especially the optimization of on-chip trace buffers) to maximize the amount of debugging information that could be provided to the user [3], [4] while minimizing overhead. In this paper, we take a step back and focus on the *user experience*. We discuss what sort of debug support we believe will make debugging HLS-generated circuits feasible for software developers, and describe our debug tool in which we have encapsulated some of these ideas.

Although we focus on functional debug in this paper, many of the ideas also apply to optimization (performance debug). The underlying challenges we will describe apply equally to both performance and functional debug, and our tool could be extended to support performance debug.

## II. The Need for Hardware Debug

As shown in Figure 1, bugs can be classified into several categories, each of which can be addressed using a different debug flow. First, *kernel-level bugs* are errors in the algorithm specification (for example, errors in loop bounds, incorrect functions, or algorithmic errors). These bugs are typically confined to one module, and are often easy to reproduce (since, often, these bugs lead to incorrect behaviour every time the circuit is run). Often, these bugs can be identified by porting, compiling, and running the original C code directly on a workstation; mapping to hardware is not necessary. In tracking down these kinds of bugs, the designer can use software debug tools (eg. gdb, Eclipse) that are already familiar to software designers.

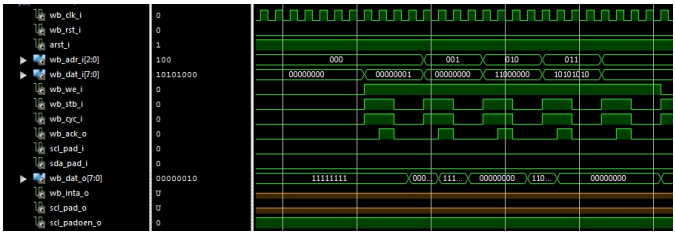

Fig. 2. Hardware view of a debug trace: Difficult for a software designer to understand!

A second class of bugs are those that appear in the generated RTL, even though the C code is correct. This may be caused by errors in the HLS tool itself, or errors in how the HLS tool is used. FPGA vendors provide the ability to uncover these bugs using a co-simulation approach where the C and RTL code is simulated on a workstation [5]. Even if the code is correct, this level of system verification is essential; until HLS tools are fully mature, many designers will appreciate the confidence they achieve from a successful RTL simulation.

Despite extensive kernel-level and RTL simulation-level testing, there will always be some design errors that escape to the hardware implementation. There are at least three reasons this can happen. First, the software emulation will run much slower than the target hardware (typically 20 to 200 times slower [6]–[9]), limiting the thoroughness of tests that can be performed. In a complex system, it is impossible to completely test (or even enumerate) all corner cases. Second, this higher-level testing will not uncover problems related to interactions with the environment or with other modules in the system, yet this is where we expect many bugs to occur. In many large systems, HLS blocks are interfaced to legacy blocks designed using RTL techniques; these interfaces may be misunderstood by the software designer leading to subtle errors that are difficult to track down. Third, the environment in which the HLS block is used (for example, the input data stream) may not be exactly as assumed during RTL simulation; inaccuracies in the model of the environment may lead to bugs that only show up when the block is connected to the real environment. For these reasons, we expect that many errors will escape simulation to the hardware design, and the only way to find these errors is to debug the system *in-situ*, running on an FPGA.

### III. Challenges for Software Engineers

The previous section made the case that certain types of design errors can only be uncovered by running a hardware implementation of the design. Debugging at the hardware level, however, is difficult for a software designer. The primary challenge during debugging an executing hardware design is that of visibility; finding the root cause of observed incorrect behaviour requires an understanding of the internal operation of the system. However, while a system is running, only I/O ports can be observed; internal signals, which are likely to provide a lot more useful information, can not be directly observed. To address this, there are commercial tools such as ChipScope from Xilinx [10], SignalTap II from Altera [11], and Certus from Mentor graphics [12]. These tools record selected signals in on-chip memories (called *trace buffers*) during the execution of the chip; at the end of the run, these trace buffers can be interrogated, and the user can use this information to understand the operation of the design, and eventually uncover the root cause of incorrect behaviour.

The challenge with this approach is that these tools provide visibility that has meaning only in the context of the generated RTL hardware. A software designer typically would not have an understanding of the underlying hardware; in fact, this is the primary reason that HLS methodologies are able to deliver high design productivity. A software designer views a design as a set of functions, each consisting of sequential control-flow code, while the underlying hardware consists of dataflow components operating in parallel across multiple clock cycles. Figure 2 shows a screenshot of the output of one of these tools; in this example, the behaviour of signals is illustrated using waveforms, a concept likely unfamiliar to many software designers. Even if understanding a waveform diagram is not a barrier, there may not be a one-to-one mapping between signals in the waveform and variables in the original C code. Further, since the HLS tool typically reorders instructions and extracts fine-grained parallelism, it is often difficult for a software designer to recognize the order of events and relate them to the order of instructions in the original C code. All of these factors make it very difficult for a software designer to use these hardware-oriented tools.

### IV. Previous Work

To our knowledge, there have been two other debugging tools produced for HLS circuits that allow source-level, in-system debug. The first such tool was presented in 2003 [13], and was designed to work in conjunction with the Sea Cucumber HLS tool [14]. However, the debugger and HLS tool both utilized the now obsolete JHDL framework and are no longer supported or available for download. The debugger allowed the user to step through the source code while the circuit running on the FPGA was executed one cycle at a time. It supported inserting breakpoints and inspecting source-code variables. Our debugger builds on some of the ideas from this work, and includes several new approaches, as explained in the next section.

The second debugging tool produced was the Inspect debugger [15] in 2014, and was designed at the same time as our work presented in [3] and [4]. Since that time we have worked with the authors to combine the ideas from their work with ours into a single tool, which will be included in the next release of the LegUp high-level synthesis tool [16].

Other work presented in [17], [18] has focused on adding debug instrumentation to HLS circuits but does not include a debugger tool.

### V. Our Debug Framework: HLS Scope

In this section, we describe how we are addressing the challenges faced by software designers debugging HLS circuits. We make our exposition concrete by presenting details of the debug tool we have created, and relate each of our ideas into features of this tool. Through this discussion, we will show that it is indeed possible to create a tool that resembles a software debugger, yet can be used to debug hardware designs running on an FPGA.

Figure 3 shows a screenshot of our tool. In the following discussions, we will refer to specific aspects of this diagram.

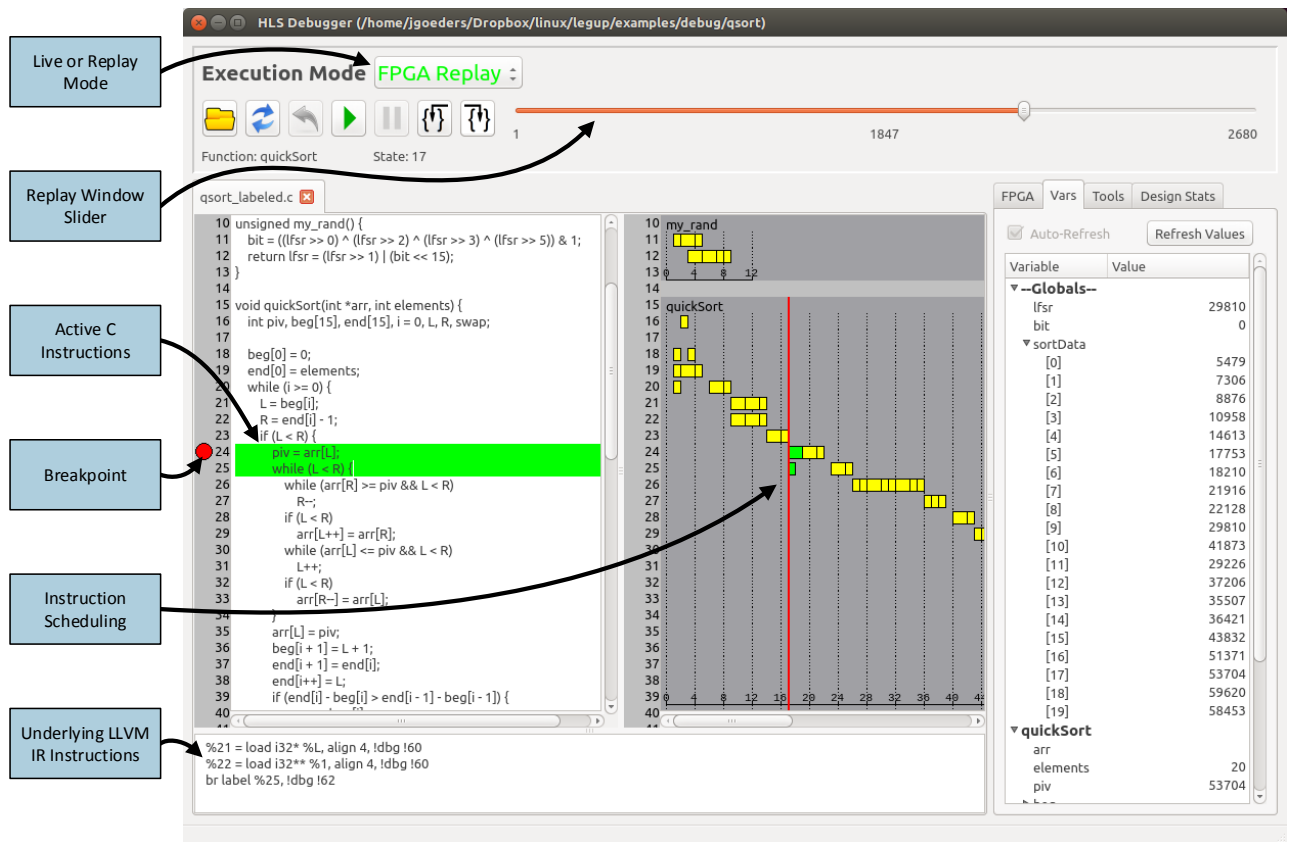

Fig. 3. Our framework: hardware debugger that a software engineer can understand

To make the debugging ecosystem easily accessible to software designers, we believe that any debug tool must resemble debug tools that these designers already understand (eg. gdb, Eclipse).[1] This means that users should be able to think of the design in terms of lines of code, and the state of the system as C-level variables. As shown in Figure 3, we achieve this in our tool; the source-level code is shown to the user (left panel), along with the C-level local and global variables (right panel).

In the left panel, the currently executing lines of source code are highlighted. The parallel nature of the circuit means that multiple C instructions can execute simultaneously. When this occurs we highlight multiple lines, as shown in Figure 3. This technique was also used in [13].

### A. Gantt Chart

As shown in Figure 3, our tool presents a Gantt chart that shows the execution of each line of code over time. This provides much of the same information as a waveform diagram, in what we believe is a more software-friendly way. The diagram provides a mechanism for the designer to understand the fine-grained parallelism that has been uncovered by the HLS tool. As an example, in Figure 3, the assignments to $beg[0]$, $end[0]$, and the initialization operation of the while loop are started during the same cycle, and this is shown graphically in the Gantt chart. Similarly, some instructions may take more than one control step (in the hardware, this corresponds to more than one clock cycle, but a clock cycle is abstracted as a control step in our tool). Long operations such as divides, or instructions that operate on arrays, typically take more than one control step; examples of the latter can be seen in Figure 3. The boxes of the Gantt chart represent individual instructions of the underlying intermediate representation (IR), which is explained further in Section V-D.

Note that in the presence of compiler optimizations, the Gantt chart may not be as straightforward as that in Figure 3; we discuss the impact of compiler optimizations in Section V-E.

### B. Debug Modes

Software designers expect to able to set breakpoints and single-step their design, and are accustomed to full visibility into the value of any variable at any point in the program. Yet, as described above, finding certain types of bugs requires running the circuit at-speed in-system. Providing enough infrastructure (trace buffers and associated logic) to provide full visibility into all signals in a hardware design running at-speed would require too much overhead to be practical. To address this, we have implemented two different debug modes: (1) *live mode*, in which the user can have full visibility, but does not run the circuit at speed, and (2) *replay mode*, in which the user can run the circuit at speed, but only has full visibility for a portion of the execution. Each of these is described below.

*1) Live Mode:* In the live mode, the system operates very much as a software debugger. The user can create breakpoints (limited by the number of hardware breakpoint units included in the instrumentation) and single step through the code. Upon hitting a breakpoint, or completing a single step, the tool disables the controlling finite state machine (FSM), essentially pausing the system. While paused, the user can retrieve the value of all variables stored in on-chip memories (right pane of Figure 3).

---
[1]Indeed, in future versions of the tool, we may investigate how we can integrate our techniques into Eclipse rather than providing a separate tool.

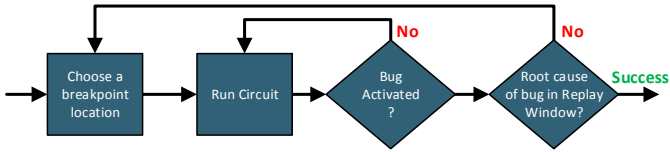

Fig. 4. Using Multiple Debug Iterations to expand the Replay Window

*2) Replay Mode:* Debugging using the Live Mode involves starting and stopping the circuit during debug, similar to the technique used in [13]. This is not conducive to running the circuit at-speed. The Replay Mode provides the ability to run the circuit at-speed while preserving a software-like debug experience. We believe that the Replay Mode is likely the most important feature in our framework and best illustrates how the software and hardware worlds can be bridged.

This mode operates as follows. While in Live Mode, the user can set a breakpoint and run the circuit to the breakpoint. As the circuit runs, instrumentation added to the circuit records the changes to signals as well as the control flow executed by the program (these values are stored on-chip in trace buffer memories). After the program hits the breakpoint, the values in the trace buffer are read into the debug tool, and the user can enter the Replay Mode. While in Replay Mode, the user can still single-step and set breakpoints as before, however, all variable values and control flow information is obtained from the trace buffer data rather than the live values from the chip. In this way, the user can observe what the chip did during the at-speed run, while maintaining a software-like debug interface. In addition to single stepping, the user can use a slider to move ahead further in the buffer, as shown in the top-right of Figure 3. Interestingly, the slider can also be used to step backwards in time, providing the illusion of running the chip backwards. We anticipate that this feature will be important as users wish to "work backwards" to determine the root cause of unexpected behaviour. In fact, the technique of working backwards is already used in the software domain; examples include gdb's *Reverse Debugging* and Microsoft Visual Studio's *IntelliTrace*.

Note that the Live Mode requires on-chip trace buffers and associated logic to be added to the circuit. Because on-chip resources are limited, we can only store data for a limited number of instructions; we refer to the length of code for which data can be stored as the *Replay Window*. Within the replay window, we can provide a complete control-flow trace, allowing the user to observe which instructions are active for each execution step. Any variables that are updated within the replay window are available for inspection, after the point they are updated. Their value is unknown prior to the first update within the replay window. To handle the case where a variable is never updated during the replay window, we include instrumentation to provide the debugger with access to the memory controllers in the circuit. This allows us to read the value of a variable directly out of memory. While in Replay Mode, the user can step forwards and backwards through all instructions in the Replay Window, but can not step outside it. If the user wishes to go outside the Replay Window, multiple debug iterations are required as shown in Figure 4.

In our previous publications [3], [4], we present efficient buffer structures as well as methods of effectively compressing control flow and data information before storing it in the buffer. Those papers show that it is possible to record, on average, control and data information for 4322 lines of C code for each 100Kb of on-chip memory used for trace buffers. We could further reduce this by only recording selected signals or using off-line reconstruction methods; however, we have not yet investigated this further.

We refer the readers to [3], [4] for further details on the debug instrumentation, including data on execution trace length, area overhead, and impact on operating frequency.

*C. Instruction-Level Parallelism*

It is important to note that there is a significant difference between single step in our framework and single step in a software debugger. Single step in a software debugger will typically advance one source code statement. A single step in our framework may advance through multiple instructions at the same time, if those instructions are mapped to hardware that execute in the same clock cycle. As an example, in Figure 3, if the system is advanced one "instruction" beyond the red vertical line, both the assignment to $piv$ and part of the $(L < R)$ check will be performed. Because this is hardware, and both are mapped to the same clock cycle, it is not possible to decouple these two operations and only single step through one of them. It may be possible to address these types of situations by creating a tool that modified the user circuit to either temporally separate these two operations or provide selective gating to each operation, however, that would result in significant changes to the user circuit, meaning the circuit being debugged may be very different than the original circuit.

In [13] the authors explored another approach. They provided virtual serialization, making it appear to the user that the statements were executed serially, when in reality they were executed in parallel in the hardware. Indeed this makes the debugger behave more like a traditional software debugger; however, this hides the parallelism from the user, and prevents them from restructuring their C code to explore different fine-grained parallelism optimizations.

*D. IR Instructions*

As shown in Figure 5, most HLS compilers compile software to hardware in several stages. First, C-to-C transforms such as loop restructuring are often performed, to make the C code more amenable to acceleration. The code is then often converted to an Intermediate Representation (IR) which resembles assembly language; several IR instructions are typically associated with each C operation. Optimizations are then

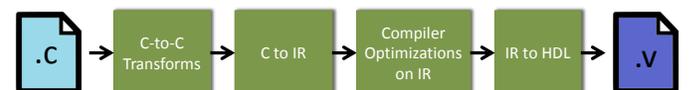

Fig. 5. Compilation flow in a typical HLS tool, illustrating the importance of the Intermediate Representation (IR)

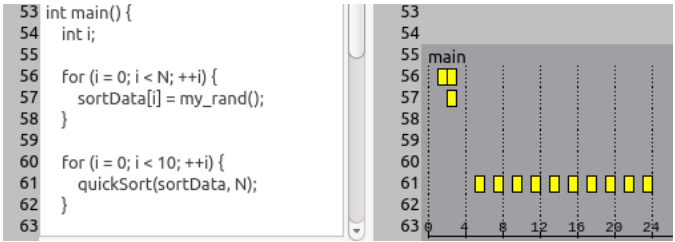

Fig. 6. Gantt chart showing loop unrolling optimization.

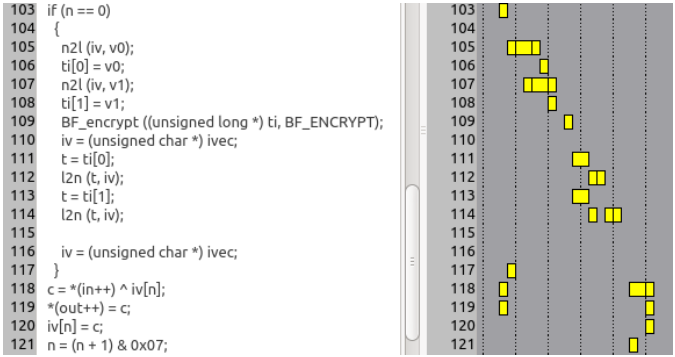

Fig. 7. Gantt chart showing code reordering optimization

performed and hardware is constructed directly from the IR representation.

Ideally, the software designer should be sheltered from the IR, and should not have to know of its existence. This is similar to how someone developing software should not have to know about the underlying assembly language of their compiled software. We anticipate that there are several reasons, however, that the HLS designer might want to inspect and understand the IR. First, C instructions often take multiple control steps, and it may sometimes become important to understand when primitive operations occur (this may be especially important when debugging multi-threaded applications). Second, during performance optimization, it may become important to understand why certain instructions take longer to execute than others; this information could be used to restructure the C code to lead to different fine-grained parallelism optimizations. Third, this information could be used by HLS tool vendors to help them as they optimize and debug their own compilers. For all these reasons, we have elected to expose the IR to the user. As shown in Figure 3, the bottom panel of the screen shows the IR for the executing instructions.

### E. Compiler Optimizations

Most HLS tools provide the ability to select the level of optimization applied to the code before hardware is generated. For HLS tools built around the LLVM framework, the user can specify the familiar -O0, -O3, etc. flags. The higher the level of optimization, the more restructuring of the code that is performed before hardware is built.

Debugging optimized code (such as generated by the -O3 flag) is notoriously difficult. Because of this, when designing software, it is common practice to debug using the -O0 (unoptimized). We believe that this strategy does *not* work well for HLS-generated circuits for two reasons. First, changing the level of optimization will significantly change the timing of the resulting circuit. Since many of the bugs we anticipate are in the interfaces between blocks, changing the timing may result in significantly different behaviours. Thus, we believe that if the "-O3" version of a circuit is going to be "shipped" it is important to debug the "-O3" version of a circuit directly. Second, in many cases, we have found that compiling with "-O0" leads to much larger circuits, some of which would not fit on the target FPGA, making on-chip debugging impossible. This is different than software, in which as long as the program fits in (virtual) memory, it can be debugged.

Because of these reasons, we have designed our tool such that it can be used to debug optimized circuits. This has two implications. First, when optimizing circuits, variables are sometimes stored as registers in the datapath rather than in local memories. This requires a change in the instrumentation (as is discussed in [4]), however, it does not necessarily impact the user experience (unless variables are optimized away all together). Second, it means that the temporal relationship between various instructions may vary greatly; the result of a single-step may not be intuitive, since several instructions that are not together in the original program are executed out-of-order or simultaneously. We believe that it is important to provide the user this level of visibility (rather than abstracting the sequencing to program order) since understanding the *actual* order of execution of instructions may be very important when debugging block-level interfaces or other timing-related problems.

Figures 6 and 7 provide examples of how the Gantt chart aids users in debugging optimized code. In Figure 6 a sorting operation is called 10 times in a loop. The optimizing compiler has completely unrolled the loop and replaced the code with 10 subsequent calls to the function. In Figure 7, the compiler has performed code reordering optimizations. In this case, the two instructions immediately after the *if* block are independent of the contents of the *if* block and are executed before it. In both cases the Gantt chart helps the user in understanding the optimization.

## VI. EXTENDING OUR FRAMEWORK

Although our debugging tool has been designed for use with the LegUp tool, we have designed it in a modular fashion, such that it can be extended to support other HLS tools, or expanded to explore new techniques for debugging HLS circuits. Figure 8 provides a diagram of the software organization of our tool.

At the heart of our tool is the *Debug Manager* which coordinates the debugging session. It tracks the current state of the design, and provides an API to control and observe the design. It generates signals when events occur in the circuit, such as when the state of the circuit changes or a breakpoint is encountered. The debug manager provides a *Backend API*, which allows for multiple backends to be added to the system to support different execution devices. For example, initially we supported a *Live* mode, which interacts with the FPGA, and a *Replay* mode, which uses the values from the trace buffers (both were previously described in this paper). Using the backend API we were able to very easily add a third execution mode, simulation-based execution using Modelsim. The backend API abstracts away the details of the device from the debugger tool. This abstraction could be used to test out different techniques of debug instrumentation, or circuits from different HLS tools, without needing to modify other parts of the debugger tool.

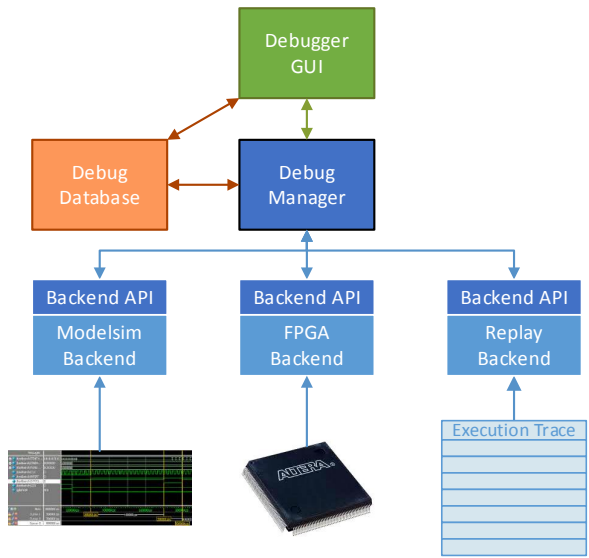

Fig. 8. Organization of our HLS-Scope debugger tool.

The *Debugger GUI*, written in Python+Qt, provides a visual overlay on top of the Debug Manager. It issues requests to the Debug Manager, such as single-stepping, obtaining the current state of the circuit, or reading variable values. The GUI could be replaced with a different tool, such as a command-line interface, binary library, or Eclipse plug-in without needing to modify the rest of the system.

The final piece of the system, and perhaps the most essential is the *Debug Database*. This is a MySQL database that contains the details of the user's design, and was adapted from that in [15]. It is automatically populated during the HLS synthesis process. It keeps track of entities in the source code (functions, lines of C code, IR instructions, variables, data types, etc.), entities in the produced circuit (modules, FSM states, signals, memories, etc.), and the relationship between them. To port our debugger to another HLS tool, it would be necessary to modify the HLS flow to populate this database.

Currently, the debugger software connects to the FPGA via UART; however, in both the instrumented hardware and the debugger tool, the UART logic is modularized, allowing it to be replaced with other communication methods.

One limitation of the current tool is that it is limited to single-threaded software; although it handles fine-grained instruction-level parallelism, it does not support coarse-grained, thread-level parallelism that is emerging in the latest HLS tools. We plan to address this in future work.

## VII. CONCLUSIONS

High-level synthesis is emerging as a mainstream design methodology, allowing software designers to target hardware implementation. As part of the HLS design process, software designers need the ability to debug their hardware systems, using debugging tools and methods familiar to them. In this paper we have presented HLS-Scope, our source-level debugger for the LegUp HLS tool. This tool is targeted to software designers, and provides a familiar debug interface, allowing them to single-step through their source code, place breakpoints and inspect variables. We include additional features such as a Gantt chart of the HLS scheduling and information of the underlying IR instructions to help bridge the gap between the sequential software and the optimized, parallelized circuit. The debug tool is open-source, modularized for extension to other applications, and available in the next release of LegUp.